\input harvmac.tex
\input epsf.tex
\overfullrule=0mm
\newcount\figno
\figno=0
\def\fig#1#2#3{
\par\begingroup\parindent=0pt\leftskip=1cm\rightskip=1cm\parindent=0pt
\baselineskip=11pt
\global\advance\figno by 1
\midinsert
\epsfxsize=#3
\centerline{\epsfbox{#2}}
{\bf Fig. \the\figno:} #1\par
\endinsert\endgroup\par
}
\def\figlabel#1{\xdef#1{\the\figno}}
\def\encadremath#1{\vbox{\hrule\hbox{\vrule\kern8pt\vbox{\kern8pt
\hbox{$\displaystyle #1$}\kern8pt}
\kern8pt\vrule}\hrule}}

\def\ptvi{\vrule height 8pt depth 6pt width 0pt}
\def\ptv{\ptvi\vrule}
%
\def\tilde{\widetilde}
\def\bar{\overline}
\def\hat{\widehat}
\def\*{\star}
\def\({\left(}		
\def\){\right)}		
\def\[{\left[}		
\def\]{\right]}

%
%
\def\frac#1#2{{#1 \over #2}}

\def\half{{1 \over 2}}
\def\d{\partial}

\def\ket#1{ | #1 \rangle}

\def\2pi{\hbox{$2\pi i$}}

\def\dsl{\raise.15ex\hbox{/}\kern-.57em\partial}
\def\Dsl{\,\raise.15ex\hbox{/}\mkern-.13.5mu D}

%
%

\def\be{\beta}
\def\al{\alpha}

\def\la{\lambda}	
\def\de{\delta}		
		\def\Om{\Omega}
\def\sig{\sigma}	

%
%

\def\CS{{\cal S}}

%
\lref\Macdonald{I.G. Macdonald, {\it Orthogonal polynomial associated with
root systems}, preprint (1988).}
\lref\sa{B.D.  Simons, B.L. Altshuler, Phys.Rev. B {\bf 50}, 1102 
(1994).}
\lref\hald{F.D.M. Haldane, Phys.Rev.Lett. {\bf 60} (1988) 635.}
\lref\shastry{B.S. Shastry, Phys.Rev.Lett. {\bf 60}, 639 (1988).}
\lref\chaine{F.D.M. Haldane, Z.N.C. Ha, J.C. Talstra, D. Bernard
and V. Pasquier, Phys.Rev.Lett. {\bf 69}, 2021 (1992).}
\lref\calo{
F. Calogero, J. Math. Phys. {\bf 10}, 2191, (1969).}
\lref\suth{B. Sutherland, J. Math. Phys. {\bf 12} , 246 (1971);
{\bf 12} , 251 (1971).}
\lref\skly{E.K. Sklyanin, J.Phys.A Math.Gen {\bf 21} (1988) 2375.}
\lref\YB{D.Bernard, M.Gaudin, F.D.M.Haldane and V.Pasquier, J.Phys.A {
\bf 26} (1993) 5219.}
\lref\vincent{V. Pasquier, in preparation.}
\lref\op{M.A. Olshanetsky, A.M. Perelomov, Phys.Rep. {\bf 94} (1983)
313.}

\Title{SPhT/95/003}{
\vbox{\centerline{Exact Solution of Long-Range}
\centerline{Interacting Spin Chains with Boundaries}}}

\centerline{D. Bernard\footnote{*}{Member of the CNRS}, V. Pasquier and D. Serban}

\centerline{Service de Physique Th\'eorique
\footnote{$^\dagger$}{Laboratoire de la Direction des Sciences
de la Mati\`ere du 
Commissariat \`a l'\'Energie Atomique},}
\centerline{CE Saclay, 91191 Gif-sur-Yvette, France}

\vskip 1.5 truecm
Abstract

We consider integrable models of the Haldane-Shastry
type with open boundary conditions. We define 
monodromy matrices, obeying the reflection equation,
which generate the symmetries of these models.
Using a map to the Calogero-Sutherland Hamiltonian
of BC type, we derive the spectrum and the
highest weight eigenstates.

\Date{1/95}

It has recently been understood that while the long-range
interacting spin chains capture some of the essential physical
features of the Heisenberg spin chain, exact results
can often be obtained more explicitly.
In particular, the spectrum
can be described in terms of elementary excitations
called spinons. 
The expressions of the wave functions can be derived.
In this paper we propose generalizations of the
Haldane-Shastry Hamiltonian \hald \shastry ~
which describe open spin chains with various kinds
of boundary conditions.

\newsec{ The Hamiltonians}

We consider a system of spins on a one-dimensional semi-circular
lattice, with an interaction depending on the inverse square of the 
distance between the spins  
\eqn\ham{
H=  
\sum_{i\neq j=1}^{N} \[ h_{ij}(P_{ij}-1) 
+ \bar h_{ij}  (\bar P_{ij}-1) \]
+ \sum_{i=1}^{N} (b_1 h_{i0}+2b_2 \bar h_{ii} )(P_i-1) .
}
The operators $P_{ij}$ permute the spins at the sites $i$ and $j$. They 
correspond to a Heisenberg-type interaction,
$P_{ij}=(\sig_i^a\sig_j^a+1)/2$, with $\sigma^a$ the Pauli matrices.
$P_i$ is a reflection operator, $P_i^2=1$, whose representation will
be chosen later and $\bar P_{ij}=P_iP_j P_{ij}$. 
The coupling constants $h_{ij}$ are defined as follows:
\eqn\hij{
h_{ij}=-\frac{z_iz_j}{(z_i-z_j)^2},\quad 
\bar h_{ij}=-\frac{z_iz_j^{-1}}{(z_i-z_j^{-1})^2}, \quad
h_{i0}=-\frac{z_i}{(z_i-1)^2}
}
with $z_i$ points on the unit circle characterizing the positions 
of the spins.

 There are at least two possible representations for the reflection
operators $P_i$; the simplest one is to put $P_i=1$, which
leads to $P_{ij}=\bar P_{ij}$. The Hamiltonian
\ham\ then becomes
\eqn\alt{
H=  
\sum_{i\neq j=1}^{N} \( h_{ij} 
+ \bar h_{ij} \) (P_{ij}-1) .
}
This model is $su(2)$ symmetric. It was first considered  
by Simons and Altshuler \sa .

The reflection operators can be equally represented by 
the Pauli matrices $P_i=\sigma_i^3$, in which case the
$su(2)$ invariance is lost; only the projection of the total spin 
is conserved.

These models can be interpreted as open versions of the 
Haldane-Shastry spin model \hald\ \shastry . The terms involving $h_{ij}$ couple
the `real' spins $i$ and $j$, while the terms with coupling
constants $\bar h_{ij}$ correspond to
the interaction of the spin $i$ with the image of the spin $j$.

When $P_i=\sigma_i^3$, the last two terms in \ham\ can be interpreted as space 
dependent magnetic fields varying as $1/{\rm sin}^2 x$, with
$x$ being the angle (or two times the angle) between 
a spin and the boundary. Remark that for the term $\bar h_{ii}$
the two boundaries are equivalent, while for the term $h_{i0}$
they are not. This term breaks the $z_i\leftrightarrow -z_i$ invariance.

\newsec{Integrability Condition}

This Hamiltonian is integrable only for certain values of the
numbers $z_i$ and of the parameters $b_1$, $b_2$. To obtain 
the condition of integrability we shall follow 
the approach of ref. \YB .
In addition to the operators which permute the spins 
, let us introduce permutation and reflection operators
acting on the coordinates as
\eqn\perm{
K_{ij}~z_i=z_j~K_{ij},\quad K_i~z_i=z_i^{-1}K_i, \quad \bar K_{ij}~z_i
=z_j^{-1}\bar K_{ij}.
}
Introduce also a set of mutually commuting operators (Dunkl operators)
\vincent\
\eqn\Dunkl{
d_i=\sum_{j;j>i}\theta_{ij}K_{ij}-\sum_{j;j<i}\theta_{ji}K_{ij}+
\sum_{i\neq j} \bar \theta_{ij} \bar K_{ij}+(b_1 \theta_{i0}+
b_2 \bar \theta_{ii})K_i
}
with $\theta_{ij}=z_i/(z_i-z_j)$, $\bar \theta_{ij}=z_i/
(z_i-z_j^{-1})$ and $\theta_{i0}=z_i/(z_i-1)$.
We denote by $\pi$ a projection operation which consists in replacing the
coordinate permutations (reflections) by spin permutations
(reflections) when they are at the right of an expression.
Let us consider the Hamiltonian $\hat H$ with the property
$H=\pi (\hat H)$
\eqn\hham{
\hat H=  
\sum_{i\neq j=1}^{N} \[ h_{ij}(K_{ij}-1) 
+ \bar h_{ij}  (\bar K_{ij}-1) \]
+ \sum_{i=1}^{N}(b_1 h_{i0}+2b_2 \bar h_{ii})(K_i-1) .
}
This Hamiltonian commutes with the reflection and permutation
operators, $[\hat H, K_i]=0$, $[\hat H, K_{ij}]=0$,
so its eigenfunctions can be chosen either symmetric or 
antisymmetric under the reflection $z_i \rightarrow z_i^{-1}$
and permutation of coordinates.
$\hat H$ can be diagonalized on a basis of functions 
depending on coordinates and spins, $\Psi$. These are also the 
eigenfunctions of $H$, provided that $(K_i-P_i)\Psi=0$, $(K_{ij}-P_{ij})
\Psi=0$.
Choosing $K_i=+1$ or $-1$ lead to the choice $P_i=1$ or
$P_i=\sigma_i^3$.

The integrability condition is that the Hamiltonian \hham\
commutes with the Dunkl operators, $[\hat H,d_i ]=0$. It 
can be translated into $N$ equations for the coordinates $z_i$
and the constants $b_{1,2}$
\eqn\coord{
2\sum_{j;j\ne i}^N (h_{ij}w_{ij}+\bar h_{ij} \bar w_{ij})
+2~b_2^2~ \bar h_{ii} \bar w_{ii} +b_1(b_1+b_2)~h_{i0}w_{i0}=0
}
where $w_{ij}=(z_i+z_j)/(z_i-z_j)$, $\bar w_{ij}=(z_i+z_j^{-1})/
(z_i-z_j^{-1})$ and  $w_{i0}=(z_i+1)/(z_i-1)$.
The system of equations $\sum_{j;j\ne i}^L
h_{ij}w_{ij}=0$ has the solution $z_k=C{\rm e}^{2i\pi k/L}$ with
$C$ a multiplicative constant. We derive solutions for the 
equations \coord\ by putting them in this form. 
Different solutions correspond to different 
distributions of the coordinates $z_i$, as illustrated in 
Fig. 1.
\fig{Distribution of the spins on the lattice. The bold line represents
the boundary (mirror), the filled circles indicate the positions of
the spins. The open dots are the images of the 'physical'
points through the mirror.}{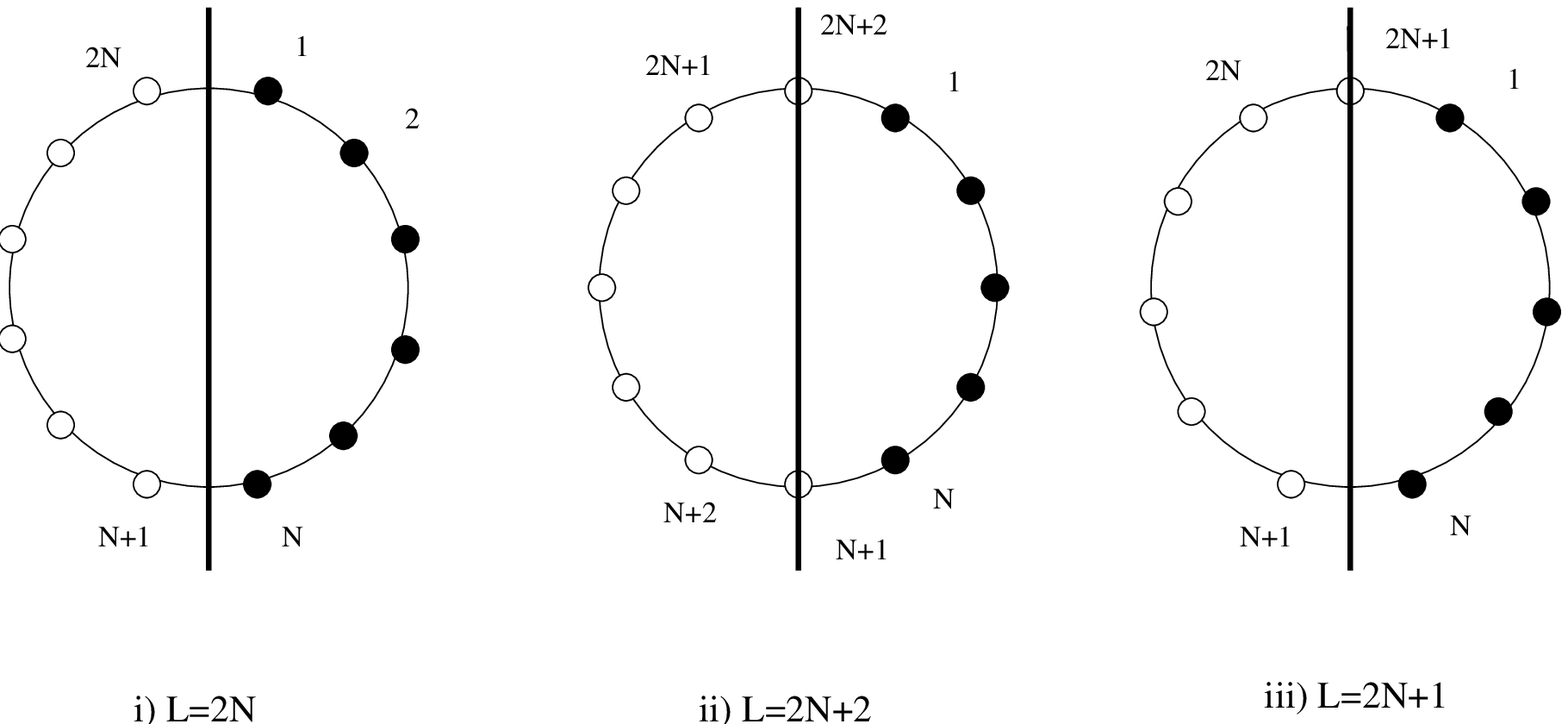}{12cm}
\figlabel\tabb
We consider a circular lattice with $L$ sites and set 
$\omega={\rm e}^{2i\pi/L}$ :
  
$i$) If the `mirror'
(the boundary) is placed between the sites of the lattice, then
$L=2N$, $N$ being the number of `physical' sites. In this
case, $z_k=\omega^{k-1/2}$ and the constants in \ham\ have the values $b_1=0$, $b_2=\pm 1$.

$ii$) If the mirror passes through the sites of the lattice, 
$L=2N+2$ and $z_k=\omega^k$. The constants in the Hamiltonian
have the values $b_1=0$, $b_2=\pm3$.

$iii$) The third case corresponds to a mirror passing through a site 
and a bond, so $L$ is odd, $L=2N+1$. The positions of the sites
are $z_k=\omega^k$ and $b_2=\pm1$. The constant $b_1$ acquires a 
non-zero value and it is the solution of the equation
$b_1(b_1+b_2)=2$.

When the reflection operator is represented by the identity,
the constants $b_1$ and $b_2$ are not relevant,
as they multiply constant terms in \ham . 
When the reflection is represented by
$\sigma_i^3$, an overall change of sign of the constants $b_1$, $b_2$ 
corresponds to the conjugation
$\sigma_i^3 \rightarrow - \sigma_i^3$, $\sigma_i^{\pm} \rightarrow 
\sigma_i^{\mp}$. In this case we will retain only the positive value
of $b_2$.

\newsec{Symmetry and Spectrum}

The model possesses a monodromy
matrix which commutes with the Hamiltonian $H$ and which 
satisfies the reflection equation \skly ~:
\eqn\rtrt{
R_{00'}(u-v)T^0(u) \bar R_{00'}(u+v)T^{0'}(v)=
T^{0'}(v) \bar R_{00'}(u+v) T^0(u) R_{00'}(u-v) 
}
where $T^0(u)$ stands for $T(u)\otimes1$ and $T^{0'}(u)$ for 
$1 \otimes T(u)$. The matrices $R(u)$ and $\bar R(u)$ are given by~:
\eqn\matr{
R(u)=u+P_{00'}, \qquad \bar R(u)=u+\bar P_{00'}
}
where the operator $P_{00'}$ permutes the two auxiliary spaces $0$
and $0'$ and $\bar P_{00'}$ represents permutation followed by a 
reflection in each space. 
The explicit expression of this monodromy matrix is 
\eqn\trans{
T(u)=\pi\[ \prod_{i=1}^N \(1+\frac{P_{i0}}{u-d_i}\)
\(1+\frac{b_1+b_2}{2} \frac{P_0}{u}\) \prod_{i=N}^1 \(1+\frac{\bar P_{i0}}{u+d_i}\)\]. 
}

Expanding $T(u)$ in powers of $u^{-1}$ around the point at the infinity
generates conserved charges, $T(u)=I+\sum_{n\geq 0} T_n u^{-(n+1)}$.
If $P_i=1$, the first conserved charge is the total spin $Q_0^a=
\sum_{i=1}^N \sigma_i^a$ and the non-trivial charges are located at 
even levels, $Q_{2n}$. 
If $P_i=\sigma_i^3$, the conserved charge at the level zero
is the projection of
the total spin, $Q_0^3=\sum_{i=1}^N \sigma_i^3$. 
The $su(2)$ invariance is broken by the presence of the last two terms in the 
Hamiltonian \ham . At the next level, we find two non-trivial
generators~:
\eqn\yang{
Q_1^{\pm}=\half \sum_{i=1}^N \sigma_i^{\pm}
\[ \sum_{j;j\neq i} \(w_{ij}P_{ij}+\bar w_{ij} \bar P_{ij} \)+
b_1 w_{i0} P_i +b_2 \bar w_{ii} P_i\].
}

As a consequence of these symmetries, the energy 
levels are degenerate and the eigenvectors are grouped in multiplets.
Each multiplet is characterized by a set of rapidities $\{m_i\}$, 
$c_1/2+c_2\leq m_i \leq L/2-1$, where $c_1$, $c_2$ and $L/2-m_i$
are integers. The values of $c_1$ and $c_2$ corresponding to
the different cases are summarized below
$$\vbox{\offinterlineskip
\halign{\ptv\quad # &\quad \ptv \quad # & \quad\ptv \quad 
# & \quad \ptv \qquad  # &  \quad \ptv #\cr 
\noalign{\hrule}
\ptvi $\ $&$\ $&$ P_i=1 $& $ P_i=\sigma_i^3 $& \cr
\noalign{\hrule}
\ptvi $L$ &$m_i$  &$c_1 \quad c_2$ & $\ b_1 \quad b_2 
\qquad c_1 \quad c_2 $ &\cr
\noalign{\hrule}
\ptvi $2N$ &$Z$  &$0\quad \ 1$ & $\ 0 \quad \ 1 
\qquad 0 \quad \ 0$ &\cr
\ptvi $2N+2$ &$Z$  &$0 \quad \ 2$ & $\ 0 \quad 
\  3 \qquad 0 \quad \ 1$ &\cr
\ptvi $2N+1$ &$Z+1/2$ &$1 \quad \ 1$ & $\ 1 \quad 
\ 1 \qquad 1 \quad \ 0$ &\cr
\ptvi $\ $ &$\ $ &$\ $ & $-2 \quad \ 1 \qquad 3 \quad \ 0$ &\cr
\noalign{\hrule} }} $$
\centerline{Table 1.}

 The energy of a
multiplet have the expression
\eqn\energ{
E_{\{m_i\}}=\sum_{j=1}^M \(m_j^2-\frac{L^2}{4}\). 
}
The form of the spectrum will be justified later, using a map to 
a continuous model, the Calogero-Sutherland model.

The rules giving the degeneracies of the spectrum depend on the symmetry of
the model, hence on the representation chosen for the operators $P_i$.
They have been determined numerically and
they should also follow from a careful analysis of the representation 
of the symmetry algebra \rtrt .
If $P_i=1$, the model possesses a $su(2)$ symmetry and the degeneracies
are the same as for the Haldane-Shastry model \chaine 
. The only difference is that 
the energy is not invariant under the transformation $m_i \rightarrow N-m_i$.
The rule can be formulated as follows: take a chain of $N$ 
fictitious spins with 
values $+$ and $-$ (these {\it are not} the spins in \ham ). Put 
a $1$ between two consecutive fictitious 
spins if the first one is greater than 
the second and a $0$ otherwise. This sequence of $0$ and $1$ is 
called a `motif' and the positions on the motif are labeled by 
integers (half-integers) ranging from $c_1/2+c_2$ to $L/2-1$.
The positions of the $1$'s give the set of quantum numbers $\{m_i\}$.
The degeneracy associated to a given motif equals the number of 
configurations of the fictitious spins compatible with it.

If $P_i=\sigma_i^3$, the symmetry group is $u(1)$  and we expect the
spectrum to be less degenerate.
In this case, the rule giving the degeneracies is
slightly modified from the one above. We must consider
now a chain of $N+2$ fictitious spins, with the first and
the last ones fixed at the value $+$. This constraint
imposes to all the fictitious spins at the left of the first $1$
to have the value $+$. The positions of the symbols 
on the motif range from $c_1/2+c_2$ to $L/2$. 
For the case $L=2N+1$,
$b_1=-2$ we obtained numerical evidence for a $su(2)$ symmetry,
but we did not succeed
to find the conserved generators (which are not the total spin operators).
In this case, the degeneracies are the same as for the 
periodic chain.

\newsec{Mapping to the Calogero-Sutherand Model} 

The eigenvectors and the eigenvalues of the Hamiltonian \ham\
can be obtained from a correspondence with a particular model
of Calogero-Sutherland (CS) type.
Before giving the details of the correspondence, let us give the
definition of the general CS model.
This model describes $n$ particles on a line, with coordinates
$x_i$, $0\leq x_i \leq l$. In the version proposed by Calogero 
and Sutherland \calo\
\suth, the interaction among the particles is pairwise
and the model is completely integrable. In fact, the model is integrable for
more general potentials, associated to all the root systems of the 
simple Lie algebras \op . 

Let $V$ denote a $n$ dimensional vector space with an orthonormal 
basis $\{e_1,...
,e_n\}$ and let $R=\{\al \}$ be a root system spanning $V$, 
with $R_+$ the set of positive roots. Let $x$ denote the
vector $(x_1,...,x_n)$ and $x\cdot \al$ its scalar product with 
the vector $\al$. Then the general CS Hamiltonian is
\eqn\CS{
H_{CS}=-\sum_{i=1}^n \frac{\d^2}{\d x_i^2}+
\sum_{\al\in R_+} \frac {g_{\al}}{{\rm sin}^2 (x\cdot\al)}
}
where $g_{\al}$ is constant on each orbit of the Weyl group (it has the
same value for the roots of the same length).
The Hamiltonian proposed by Sutherland and used to solve
the Haldane-Shastry model correspond to the 
root system of type $A_n$, $R=\{e_i-e_j, i\neq j \}$. The one relevant 
for our problem is associated to
the $BC_n$ (generalized) root system,
$R=\{\pm e_i,\ \pm 2e_i,\ \pm e_i \pm e_j\}$. In this case,
the Weyl group is generated by the $n$ reflections on the hyperplanes
$x_i=0$ ($K_i$) and by the reflections on the hyperplanes $x_i=\pm x_j$
(the permutations $K_{ij}$, $\bar K_{ij}$).

With the variables $z_k={\rm e}^{ix_k}$
the Hamiltonian \CS\ becomes:
\eqn\calo{
H_{CS}=\sum_{i=1}^n (z_i\d_{z_i})^2 + \be(\be-1)\sum_{i\neq j=1}^n
\( h_{ij} + \bar h_{ij} \) + \sum_{i=1}^n 
\(c_1(2c_2+c_1-1) h_{i0}+ 4c_2(c_2-1) \bar h_{ii}\).
}
with $h_{ij}$ defined in terms of $z$ as in \hij . The form of the 
three independent coupling constants $g_\al$ was chosen for later convenience.
The ground state wave function of this Hamiltonian is 
\eqn\fdt{
\phi_0(z)=\prod_i z_i^{-(\be(n-1)+c_2+c_1/2)}(z_i-1)^{c_1}(z_i^2-1)^{c_2}
\prod_{i\leq j}(z_i-z_j)^\be (z_iz_j-1)^\be
}
In some situations, a gauge transformed version of \calo, $\tilde H_{CS}=
\phi_0^{-1} H_{CS} \phi_0$, can be useful
\eqn\gauge{
\tilde H_{CS}=\sum_{i=1}^n (z_i \d_{z_i})^2 +\beta \sum_{i\neq j}
(w_{ij}+\bar w_{ij})z_i \d_{z_i} + \sum_{i=1}^n (c_1w_{i0}+2c_2\bar w_{ii})
z_i\d_{z_i}+E_0
}
with $E_0=\sum_{i-1}^n\[c_1/2+c_2+\be(n-i)\]^2$.
As in the case initially studied by Sutherland, this Hamiltonian is
triangular in a basis of symmetrized plane waves
\eqn\em{
m_\la=\sum z_1^{\la_1}z_2^{\la_2}...z_n^{\la_n}
}
where the sum is now over all the permutations and the changes of sign
of the integers $\la_i$ $(\la_1\geq ...\geq \la_n \geq 0)$.

Therefore, the eigenfunctions of the Hamiltonian $H_{CS}$ can be taken
of the form
\eqn\state{
\phi(z)=\phi_0(z)P_{\la_1,...\la_n}(z)
}
where the $P(z)$ are
polynomials in the variables $z_i,z_i^{-1}$, symmetric
under the permutations $z_i \leftrightarrow z_j$ and under the
reflections $z_i \rightarrow z_i^{-1}$.
Some properties of these polynomials associated to general 
root systems are studied by Macdonald \Macdonald.
The corresponding eigenvalues are
\eqn\en{
E_{CS}=\sum_{i=1}^n(\la_i+\be (n-i)+c_1/2+c_2)^2
}

The spectrum and the eigenvectors 
of the spin chain can be derived from 
the spectrum and the wave functions of the CS model .
We construct the Hilbert space of the chain  
from the ferromagnetic state $\ket{\Om}=\ket{++\ldots +}$ by 
reversing $M$ spins~:
\eqn\etat{
\ket{\Psi}=\sum_{n_1,...,n_M=1}^{N} \psi_{n_1,...,n_M}
\sig^-_{n_1}...\sig^-_{n_M} \ket{\Om}
}
The coefficients $\psi_{n_1,...,n_M}$ are symmetric 
under the permutation of their indices.
We extend their definition for $1\leq n\leq L$ by
putting $\psi_{...,L-n_i+2\al,...}=K_i~ \psi_{...,n_i,...} = \pm 
\psi_{...,n_i,...}$ with $\al=1/2$ if $L=2N$ and $\al=0$
otherwise. The $+$ sign corresponds to the choice $P_i=1$
and the $-$ sign to $P_i=\sigma_i^3$.
By convention, we chose the coefficients equal to zero
if an index equals $L$ or if two indices coincide.

Let now $\Psi(z_1,...,z_M)$ be a symmetric polynomial of
degree at most $L-1$ in each $z_i$ such that
$\psi_{n_1,...,n_M}= \Psi(\omega^{n_1-\al },
...,\omega^{n_M-\al })$. The condition of antisymmetry of $\psi$
(when $P_i=\sigma_i^3$) and the condition for $\ket{\Psi}$ to be 
a $su(2)$ highest weight, $\sum_{i=1}^N \sigma^+_i \ket{\Psi}=0$,
(when $P_i=1$) impose that $\sum_{n_1=1}^L \psi_{n_1,...,n_M}=0$.
This means that $\Psi(z_1=0,z_2,...,z_M)=0$ and the degree
of $\Psi(z)$ in each variable is greater than $0$.

We want to translate the action of the spin Hamiltonian
on the vectors $\ket{\Psi}$ into an action on the functions $\Psi(z)$.
To do that, we use the fact that in the basis of polynomials $Q_k$ in one variable 
of degree $(L-1)$ specified by $Q_k(\omega^{n-\al})=\de_k^n$, the matrix elements of the derivatives are:
\eqn\deriv{\eqalign{
&(z\d_z-\frac{L-1}{2})Q_k(z)=-\sum_{j;j\neq k}^L \frac{\omega^k}
{\omega^k-\omega^j}Q_j(z) \cr
&z\d_z(z\d_z-L)Q_k(z)=-2\sum_{j;j\neq k}^L \frac{\omega^k\omega^j}
{(\omega^k-\omega^j)^2}(Q_j(z)-Q_k(z)) \cr
 }}
This enables to recover the action
of the Calogero-Sutherland Hamiltonian on the functions $\phi(z)=
\prod_{i=1}^Mz_i^{-L/2} \Psi(z)$~:
\eqn\corresp
{
H\ket{\Psi}=\sum_{n_1,...,n_M} (-1)^{\sum n_i}  \[ \(H_{CS}-\frac{
ML^2}{4}\) \phi \]
(\omega^{n_1-\al },...,\omega^{n_M-\al }) \sig_{n_1}^-,...,\sig_{n_M}^- 
\ket{\Om}
}
The values of the coupling constants $c_{1,2}$
in $H_{CS}$ are the values given in Table 1
and the constant $\be$ is equal to $2$. Note that, 
for these values, the energy in \en\ is consistent with \energ .
The rapidities $m_i$ are defined in terms of the integers $\la_i$
as $m_i=\la_i+2(M-i)+c_1/2+c_2$. They are integers or half-integers,
depending on the parity of $c_1$, and they obey the 
following selection rules~:
$m_i-m_{i+1}\geq 2$ and $c_1/2+c_2 \leq m_i \leq L/2-1$.
The bounds on the $m_i$ follow from the fact that
$\Psi(z)$ is a polynomial of degree between $1$ and $L-1$
in each variable $z_i$.
 
Thus, we have obtained a set of 
eigenfunctions of the spin chain from the
wave functions \state\ of the Calogero-Sutherland model. 
This procedure gives only the highest weight
vector (annihilated by $Q_0^+$ or $Q_1^+$) in each multiplet.
The other vectors 
in the multiplet can be obtained by repeated action of the 
generators $Q^a_{0,1}$.

\newsec{Acknowledgment}

We wish to thank S.K. Yang for informing us about the reference \sa .

\listrefs

\end